# Quantitative assessment of linear noise-reduction filters for spectroscopy

L. V. Le[1,2], Y. D. Kim[1,*], and D. E. Aspnes[3,*]

[1]*Department of Physics, Kyung Hee University, Seoul 02447, Republic of Korea*

[2]*Institute of Materials Science, Vietnam Academy of Science and Technology, Hanoi 100000, Vietnam*

[3]*Department of Physics, North Carolina State University, Raleigh, NC 27695-8202 USA*

*E-mail addresses: aspnes@ncsu.edu (D. E. Aspnes), ydkim@khu.ac.kr (Y. D. Kim)

**Abstract**: Linear noise-reduction filters used in spectroscopy must strike a balance between reducing noise and preserving lineshapes, the two conflicting requirements of interest. Here, we quantify this tradeoff by capitalizing on Parseval's Theorem to cast two measures of performance, mean-square error (MSE) and noise, into reciprocal- (Fourier-) space (RS). The resulting expressions are simpler and more informative than those based in direct- (spectral) space (DS). These results provide quantitative insight not only into the effectiveness of different linear filters, but also information as to how they can be improved. Surprisingly, the rectangular ("ideal" or "brick wall") filter is found to be nearly optimal, a consequence of eliminating distortion in low-order Fourier coefficients where the major fraction of spectral information is contained. Using the information provided by the RS version of MSE, we develop a version that is demonstrably superior to the brick-wall and also the Gauss-Hermite filter, its former nearest competitor.

## 1. Introduction

Reducing noise in spectra, optical or otherwise, is easy; reducing noise without compromising information is not. The challenge is particularly acute in optical spectroscopy, where the information is contained in lineshapes and the poles that give rise to features in them. While in principle noise can be reduced by taking better data, this is not always possible in practice. Given the significant advantages of working with clean spectra, noise-reduction procedures are not likely to be abandoned any time soon.

An enormous number of procedures have been developed to meet this need [1-18]. A major reason for this proliferation of approaches is the absence of an adequate method of assessing them quantitatively. A common measure is the mean-square error (MSE), which is defined as the square of the difference between the filtered lineshape and the data, summed over the spectral range [19–21]. As this involves (nominally) small differences between two relatively large sets of relatively large numbers projected onto a single value, a direct-space (DS) calculation makes the origins of these differences difficult to identify. The situation is aggravated in practice because the lineshapes themselves are generally unknown.

Here, we show that significant insight can be obtained by taking advantage of Parseval's (the Power) Theorem to cast these calculations into reciprocal- (Fourier-) space (RS). It has long been appreciated that intelligent filtering can only be done by considering the behavior of data in RS [2,3,8,9]. We show that the same advantages are realized with performance measures as well. Parseval's Theorem states that the sum of the absolute squares of a function defined at discrete, equally spaced points is equal to the sum of the absolute squares of its Fourier coefficients. Being mean-square quantities, both MSEs and noise satisfy the Parseval condition. When combined with the convolution theorem, the RS version of the MSE calculation reduces to a simple expression that shows not only how MSE errors arise, but also how they are apportioned between approximations and noise, both of which can be estimated accurately without knowing the lineshape itself. Taking advantage of the insight provided, we design a linear filter that outperforms the previous best, the extended-Gauss [9] or Gauss-Hermite (GH) filter [7], with a computational load that is significantly lower.

Conceptually, linear noise-reduction filters capitalize on the fact that information enters as point-to-point correlations and noise as point-to-point fluctuations. Thus in RS information and noise are essentially separated into low- and high-order Fourier coefficients, respectively. Nominally, the perfect linear filter multiplies all information-dominated coefficients by 1 and all noise-dominated coefficients by zero [22,23]. This rectangular, or so-called "ideal" or "brick-wall" (BW) filter, is rarely used, partly because its DS Fourier transform is wide, inhibiting its use in convolution, and partly because the Gibbs oscillations [24] resulting from its abrupt cutoff in RS compromise reconstructed lineshapes. As a result, efforts have been directed nearly universally to compromises such as the binary [12], Savitzky-Golay (SG) [15], and above-mentioned GH [7] filters, to name a few. These approximate the ideal filter via the Butterworth approach [25], i.e., eliminate as many derivatives as practical in a Taylor-series expansion of the transfer function about the lowest-index coefficient $C_0$, followed by a rolloff to zero near the white-noise cutoff $C_{n_c}$.

The results presented here illustrate the problem with these filters. The transfer functions mentioned above drop below 1 well before $n_c$ is reached, thereby compromising information. This is particularly damaging because data coefficients tend to decrease either exponentially or as Gaussians up to the white-noise cutoff. From this perspective, it may not be surprising to learn that the variation of the brick-wall filter that we present here, which is obtained by working from the high-index end, is better than any other linear filter proposed to date. Further, a parallel calculation that also takes advantage of Parseval's Theorem shows that the pass-through rms noise is only about 10% greater than that of the running-average (RA) filter, which being rectangular in DS has the lowest noise of any filter but severely distorts lineshapes. Our purposes here are to develop the theory and to provide examples.

## 2. Theory

We start by establishing the scope of the work. Continuum and discrete approaches represent two different but related methods of analysis. The continuum approach is more convenient mathematically, since it deals with integrations instead of summations, and leads to simpler analytic expressions. It also avoids complications due to the constraint of periodicity. However, data are discrete, and with procedures [26,27] available for suppressing endpoint-discontinuity artifacts that can mask information in standard Fourier analysis, the digital approach is more relevant for application. However, our objectives here are to assess methods. Consequently, we work primarily with the continuum, leaving discrete analysis for a following paper. Continuum analysis also allows direct comparisons between the two "extreme" cases, the RA and BW filters, because their continuum versions allow their RS ranges to be adjusted so their DS cutoffs are identical. This is necessary for quantitative comparisons.

Proceeding, we assume that the data are represented by a square-integrable continuous function $f(x)$. Then

$$\int_{-\infty}^{\infty} dx f(x) f^*(x) = \int_{-\infty}^{\infty} dx \, |f(x)|^2 . \tag{1}$$

is finite. Next, we define the Fourier transform $F(k)$ of $f(x)$ in the usual manner as

$$f(x) = \int_{-\infty}^{\infty} dk \, F(k) e^{ikx} , \tag{2}$$

Parseval's Theorem, discussed below, ensures that $F(k)$ is square integrable if $f(x)$ is square integrable. Now using

$$\int_{-\infty}^{\infty} dx\, e^{i(k-k')x} = 2\pi\delta(k-k'), \tag{3}$$

it follows that

$$F(k) = \frac{1}{2\pi} \int_{-\infty}^{\infty} dx\, f(x) e^{-ikx}. \tag{4}$$

Next, define the filtered (noise-averaged) function $\bar{f}(x)$ as

$$\bar{f}(x) = \int_{-\infty}^{\infty} dx'\, f(x-x') b(x'), \tag{5}$$

where the filter function $b(x)$ is unitary, that is,

$$\int_{-\infty}^{\infty} dx\, b(x) = 1. \tag{6}$$

Given Eq. (6), we define the corresponding transfer function $B(k)$ of $b(x)$ as

$$B(k) = \int_{-\infty}^{\infty} dx\, b(x) e^{-ikx}, \tag{7}$$

which ensures that $B(0) = 1$, as required for a low-pass filter. The respective inverse transform is therefore

$$b(x) = \frac{1}{2\pi} \int_{-\infty}^{\infty} dk\, B(k) e^{ikx}. \tag{8}$$

The convolution theorem is needed next. Consistent with $F(k)$, define

$$\bar{F}(k) = \frac{1}{2\pi} \int_{-\infty}^{\infty} dx\, \bar{f}(x) e^{-ikx} = \frac{1}{2\pi} \int_{-\infty}^{\infty} dx \left( \int_{-\infty}^{\infty} dx'\, f(x-x') b(x') \right) e^{-ikx}. \tag{9}$$

Substituting the respective Fourier transforms we obtain

$$\bar{F}(k) = \frac{1}{2\pi} \int_{-\infty}^{\infty} dx \int_{-\infty}^{\infty} dx' \left\{ \int_{-\infty}^{\infty} dk'\, F(k') e^{ik'(x-x')} \right\} \left\{ \frac{1}{2\pi} \int_{-\infty}^{\infty} dk''\, B(k'') e^{ik''x'} \right\} e^{-ikx} \tag{10a}$$

$$= \frac{1}{2\pi} \int_{-\infty}^{\infty} dx \int_{-\infty}^{\infty} dk'\, F(k') B(k') e^{i(k'-k)x} \tag{10b}$$

$$= F(k) B(k). \tag{10c}$$

Thus, as is well known, the Fourier coefficients of the result of a convolution are the products of the Fourier coefficients of the convolving functions.

Finally, consider

$$\int_{-\infty}^{\infty} dx\, f(x) f^*(x) = \int_{-\infty}^{\infty} dx \left( \left( \int_{-\infty}^{\infty} dk\, F(k) e^{ikx} \right) \left( \int_{-\infty}^{\infty} dk'\, F^*(k') e^{-ik'x} \right) \right) \quad (11a)$$

$$= 2\pi \int_{-\infty}^{\infty} dk\, dk'\, \delta(k-k') F(k) F^*(k') \quad (11b)$$

$$= 2\pi \int_{-\infty}^{\infty} dk\, F(k) F^*(k). \quad (11c)$$

This is Parseval's Theorem.

We now consider the mean-square error (MSE). This is defined as

$$\delta_{MSE}^2 = \int_{-\infty}^{\infty} dx\, \Delta f(x) \Delta f^*(x) \quad (12)$$

where

$$\Delta f(x) = \overline{f}(x) - f(x). \quad (13)$$

With appropriate substitutions

$$\delta_{MSE}^2 = \int_{-\infty}^{\infty} dx\, \Delta f(x) \Delta f^*(x) = \int_{-\infty}^{\infty} dx\, \left( \left( \overline{f}(x) - f(x) \right) \left( \overline{f}^*(x) - f^*(x) \right) \right) \quad (14a)$$

$$= 2\pi \int_{-\infty}^{\infty} dk\, \Delta F(k) \Delta F^*(k) \quad (14b)$$

$$= 2\pi \int_{-\infty}^{\infty} dk\, \left( \overline{F}(k) - F(k) \right) \left( \overline{F}^*(k) - F^*(k) \right) \quad (14c)$$

$$= 2\pi \int_{-\infty}^{\infty} dk\, \left( F(k) B(k) - F(k) \right) \left( F^*(k) B^*(k) - F^*(k) \right) \quad (14d)$$

$$= 2\pi \int_{-\infty}^{\infty} dk\, \left( |F(k)|^2 |1 - B(k)|^2 \right). \quad (14e)$$

The second step uses Parseval's Theorem, and the fourth step the convolution theorem.

Equation (14e) is one of the main results of this work. It shows that $\delta^2_{MSE}$ is determined by a combination of the properties of both data and filter. With $B(k) \cong 1$ for small $k$ for a well-designed low-pass filter, one might expect that the dominant contribution to $\delta^2_{MSE}$ comes from the high-$k$ end as $B(k)$ rolls off to zero. While high-$k$ behavior is certainly a factor, for typical data $F(k)$ decreases approximately exponentially or even faster with $k$, so unless $B(k)$ is accurately 1 for low $k$, this contribution to $\delta^2_{MSE}$ can be surprisingly large. Because $F(k)$ is directly calculated from the data, accurate estimates of $\delta^2_{MSE}$, along with its dominant contributions, can be obtained even if very little is known about the DS lineshape itself.

We now consider white noise, which we assume enters as fluctuations $\delta f(x)$ of $f(x)$. Although infinities can be managed, it is easier to avoid them completely by performing the derivation in discrete form, then converting the result to the continuum at the end. Noise is introduced by replacing $f_j = f(x_j) \to f_j + \delta f_j$, then using

$$\sum_{j,j'=-N}^{N} \left(\overline{f}_j^* + \delta \overline{f}_j^*\right)\left(\overline{f}_{j'} + \delta \overline{f}_{j'}\right) = \sum_{j,j',j'',j'''=-N}^{N} \left(f_{j-j''}^* + \delta f_{j-j''}^*\right) b_{j''} \left(f_{j'-j'''} + \delta f_{j'-j'''}\right) b_{j'''}, \quad (15)$$

where $b_j = b(x_j)$. The lowest-order terms are ignored because they are of no interest in the following. Assuming that the fluctuations are random, the cross terms can be ignored as well because they vanish on ensemble averaging. We are left with

$$\sum_{j,j'=-N}^{N} \delta \overline{f}_j^* \delta \overline{f}_{j'} = \sum_{j,j',j'',j'''=-N}^{N} \delta f_{j-j''}^* b_{j''} \delta f_{j'-j'''} b_{j'''}. \quad (16)$$

In the sum on the left only the diagonal terms $j = j'$ survive, and there is only one per $j$. Hence we can replace $j'$ with $j$ and eliminate the sum over $j'$. The result is

$$\sum_{j=-N}^{N} \delta \overline{f}_j^* \delta \overline{f}_j = \sum_{j,j'',j'''=-N}^{N} \delta f_{j-j''}^* b_{j''} \delta f_{j-j'''} b_{j'''}. \quad (17)$$

Next, the same argument applied to the right side shows that $j'' = j'''$, and again there is only one such term per $j''$. We are therefore left with

$$\sum_{j=-N}^{N} |\delta \overline{f}_j|^2 = \sum_{j,j''=-N}^{N} \delta f_{j-j''}^* \delta f_{j-j''} b_{j''} b_{j''} \quad (18a)$$

$$= \sum_{j=-N}^{N} |\delta f_j|^2 \left( \sum_{j''=-N}^{N} b_{j''}^2 \right). \tag{18b}$$

where in the final line we have assumed that the $\delta f_j$ are independent of $j$. The remaining sums over $j$ are superfluous, so we find that

$$|\delta \bar{f}|^2 = |\delta f|^2 \sum_{j=-N}^{N} b_j^2, \tag{19}$$

again assuming that the same uncertainty applies to all points. As expected, with the assumption that the noise is the same everywhere, the result is independent of the data. This would obviously not be the case if the noise were a function of $x$.

The continuum equivalent of Eq. (19) is

$$|\delta \bar{f}|^2 = |\delta f|^2 \int_{-\infty}^{\infty} dx\, b^2(x), \tag{20}$$

where in this equation $|\delta f|^2$ is interpreted as uncertainty per unit $x$. That is, if $|\delta f|^2$ is multiplied by the length increment $\Delta x$ between points, the result is the digital version. By Eq. (11) we can also write this result as

$$|\delta \bar{f}|^2 = \frac{|\delta f|^2}{2\pi} \int_{-\infty}^{\infty} dk\, B^2(k), \tag{21}$$

casting the calculation into RS.

Equations (20) and (21) show that white noise enters Eq. (14e) as an additive constant:

$$\delta_{MSE}^2 = 2\pi \int_{-\infty}^{\infty} dk \left( \left( |F(k)|^2 + \frac{1}{2\pi} |\delta f|^2 \right) |1 - B(k)|^2 \right), \tag{22}$$

where for the moment we view $F(k)$ as representing the information. Thus over any range where $B(k) = 1$, noise as well as information is passed unattenuated. Because $|F(k)|^2$ typically decreases exponentially or even faster with increasing $k$, Eq. (22) shows that noise constitutes a much smaller fraction of low-index coefficients than high-index coefficients.

However, the white-noise contribution continues unabated, so white noise must eventually dominate. Thus to prevent $\delta_{MSE}^2$ from increasing without limit, a cutoff is essential. We define this noise cutoff $k = k_N$ by

$$|F(k_N)|^2 = \frac{1}{2\pi}|\delta f|^2. \tag{23}$$

That is, $k_N$ is the value of $k$ where the absolute square $|F(k)|^2$ of the information equals that of the noise $|\delta f|^2/2\pi$. In filter terms this can be approximated as

$$B(k_N) = 0.5, \tag{24}$$

noting that $B(0) = 1$. With this definition nearly all of the information is included and the avoidable noise contribution is minimized. Once $k_N$ has been determined, the RS scales of the different filters discussed below can be set accordingly. Thus for several reasons intelligent filtering requires an assessment of the data in RS.

However, when comparing the intrinsic performances of linear filters, the relevant cutoffs are in DS, not RS. Accordingly, in what follows we adjust the RS ranges of all filters such that their DS cutoffs $x = x_c$, defined as

$$b(x_c)/b(0) = 0.5, \tag{25}$$

are identical, using as reference the RA filter, as noted above. For example, this determines the RS cutoff of the BW filter discussed below.

Finally, cutoffs generate Gibbs oscillations [24]. This is especially true for high-performance linear filters. Although not a direct measure of performance, these artifacts can be estimated. Given Eqs. (2) and (5), it is seen that the "missing" contribution is given by

$$\Delta f_G(x) = \int_{-\infty}^{\infty} dk \, F(k)(1 - B(k)) \, e^{ikx}, \tag{26}$$

where $k_c$ is the continuum equivalent of $n_c$. For filters with a relatively sharp cutoff $k = k_c$, Eq. (26) shows that the magnitude and frequency of the cutoff oscillations are of the order of the magnitude and frequency of the transform of the last RS coefficient prior to cutoff.

These lineshape errors can be eliminated, or at least greatly reduced, by replacing the attenuated coefficients with their unattenuated equivalents obtained by extrapolating trends established by the low-order coefficients into the white-noise region. This can be done either by model-dependent curvefitting or model-independent maximum-entropy analysis [2]. That both procedures are nonlinear is easily proven: neither can be represented as DS convolutions. Because Gibbs oscillations are relatively easy to eliminate this way, we consider them to be a non-issue. As a result we conclude that in any given situation a combination of linear and nonlinear filtering will outperform linear filtering alone. This is the subject of a subsequent paper.

## 3. Running-average and brick-wall filters

We now consider specific linear filters, starting with RA and BW filters, which are rectangular in DS and RS, respectively, and hence represent extreme cases. To compare filters, Eq. (14e) shows that data are necessary. We represent these as a single Lorentzian (lifetime-broadened) absorption line of full-width-half-maximum $2\Gamma$, which yields relatively simple but relevant analytic solutions. Results are expressed as a ratio $\eta = \Gamma/x_o$, where $2x_o$ is the full width of the RA filter in DS.

The RA and BW filters are defined as

$$b_{RA}(x) = \frac{1}{2x_o} u(x_o - |x|); \tag{27}$$

$$B_{BW}(k) = u(k_o - |k|). \tag{28}$$

Their respective Fourier transforms are

$$B_{RA}(k) = \frac{\sin(kx_o)}{kx_o}, \tag{29}$$

$$b_{BW}(x) = \frac{\sin k_o x}{\pi x}. \tag{30}$$

In both cases $B_{RA}(0)$ and $B_{BW}(0)$ are equal to 1, as required for low-pass filters, and hence the respective $b(x)$ are unitary. As a cross-check, it is easy to show that these expressions also satisfy Parseval's Theorem. We standardize all filtering performances by choosing the RA cutoff $x_o$ in DS to be $x_o = 1$, then adjust suitable parameters in other filters ($k_o$ in the case of the BW filter) such that $b(1)$ is half the value of $b(0)$. For the BW filter this gives

$$\frac{b(x_o)}{b(0)} = 0.5 = \frac{(k_o/\pi)\text{sinc}(k_o x_o)}{(k_o/\pi)} = \text{sinc}(k_o x_o), \tag{31}$$

or

$$k_o x_o \cong 1.895. \tag{32}$$

The DS lineshapes of the two limiting filters are shown in Fig. 1, and their transfer functions in Fig. 2. As can be appreciated, the filters are (x,k) complementary in that the lineshape of $b(x)$ for the BW filter is the RA lineshape of $B(k)$, and vice versa. Also, the ratios $b(x_o)/b(0)$ and $B(k_o)/B(0)$ are equal to 1/2 in both cases.

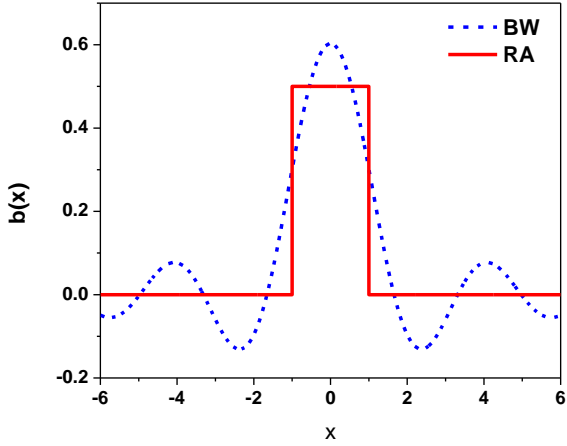
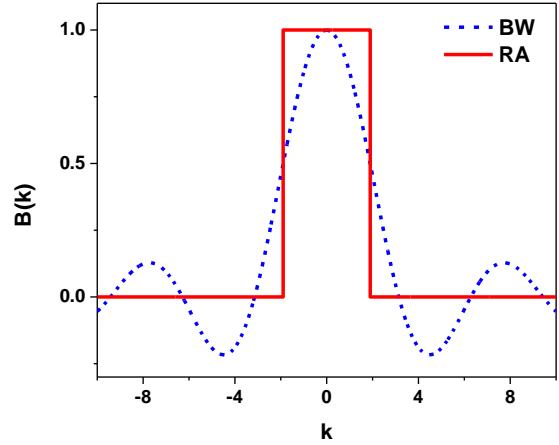

**Fig. 1.** DS convolution lineshapes $b(x)$ of the running-average (RA, red line) and "ideal" or brick-wall (BW, blue line) filters as functions of $x$ as obtained from Eqs. (27) and (30), respectively. The cutoff of both filters is $x_o = 1$.

**Fig. 2.** The transfer functions of the RA and BW filters of Fig. 1.

We next consider figures of merit, starting with noise, which for uniform noise per data point is independent of the data. Substituting the above expressions into Eqs. (20) and (21) gives the rms values

$$\sqrt{|\delta f|^2_{noise,RA}} = \frac{1}{\sqrt{2x_o}} = \frac{0.707}{\sqrt{x_o}} ; \tag{33}$$

$$\sqrt{|\delta f|^2_{noise,BW}} = \sqrt{\frac{k_o}{\pi}} = \sqrt{\frac{1.895}{\pi x_o}} = \frac{0.777}{\sqrt{x_o}} . \tag{34}$$

Thus with the same DS cutoffs, the BW filter exhibits only about 10% more noise than the RA filter. This is surprisingly low, given that $b(x)$ for the BW filter has a relatively long range in DS.

To compare MSEs, we assume as noted above that the data are represented by a Lorentzian lineshape normalized to unit area:

$$f(x) = \frac{\Gamma}{\pi} \frac{1}{x^2 + \Gamma^2}. \tag{35}$$

Its Fourier transform is

$$F(k) = \frac{1}{2\pi} e^{-|k|\Gamma}. \tag{36}$$

Given these expressions, it follows immediately that

$$\delta^2_{MSE,BW} = \frac{1}{2\pi\Gamma} e^{-2k_o\Gamma} = \frac{1}{2\pi\eta x_o} e^{-3.79\eta x_o}, \tag{37}$$

We define $\eta = \Gamma/x_o$ as the ratio of lineshape width to filter width. In practical applications where the width of the filter is significantly less than the width of the structures in a spectrum, typical values of $\eta$ range from about 2 to 10.

Using Eq. (36) we find that

$$\delta^2_{MSE,RA} = 2\pi \int_{-\infty}^{\infty} dk\, F^2(k)\left(1 - B^2(k)\right) \tag{38a}$$

$$= 2\pi \int_{-\infty}^{\infty} dk\, \frac{1}{4\pi^2} e^{-2|k|\Gamma} \left(1 - \frac{\sin kx_o}{kx_o}\right)^2 \tag{38b}$$

$$= \frac{1}{\pi x_o} \int_0^{\infty} dy\, e^{-2\eta y} \left(1 - \frac{\sin y}{y}\right)^2, \tag{38c}$$

By expanding the square, integrating the last term by parts, and taking advantage of a good table of integrals, the closed-form version of Eq. (38c) is found to be

$$\delta^2_{MSE,RA} = \frac{1}{\pi x_o} \left( \frac{1}{2\eta} - 2\tan^{-1}\frac{1}{2\eta} - \frac{\eta}{2}\ln\left(1 + \frac{1}{\eta^2}\right) + \tan^{-1}\frac{1}{\eta} \right). \tag{39}$$

For $\eta \to 0$, that is, in the limit that the Lorentzian effectively reduces to a delta function, both expressions diverge as $1/\eta$, as a consequence of peak values becoming infinite, even though their integrated areas are finite. Accordingly, it is more instructive to consider the ratio, which is

$$\delta^2_{MSE,RA} / \delta^2_{MSE,BW} = e^{3.79\eta} \left( 1 - 4\eta\tan^{-1}\frac{1}{2\eta} - \eta^2 \ln\left(1 + \frac{1}{\eta^2}\right) + 2\eta\tan^{-1}\frac{1}{\eta} \right). \tag{40}$$

This ratio is shown for $0 \leq \eta \leq 1.5$ and $4 \leq \eta \leq 5$ in Figs. 3 and 4, respectively. Figure 3 shows that the RA is initially better, but by $\eta \sim 1$ the BW starts to dominate, after which the difference increases exponentially. Figure 4 shows a more typical range encountered in practice. For these values of $\eta$ the exponential dependence has taken over completely, and the BW filter is better by orders of magnitude. Thus, of the two extreme cases, the BW result, Eq. (37), is the useful benchmark.

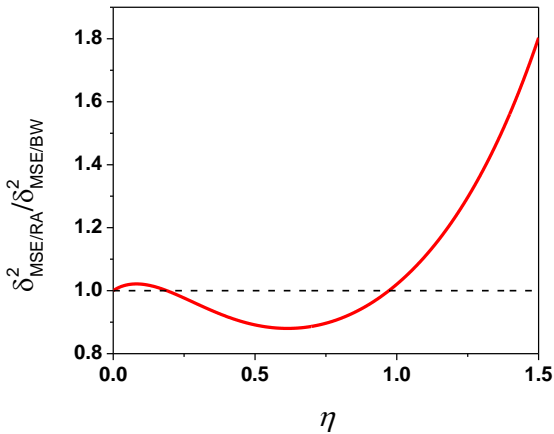
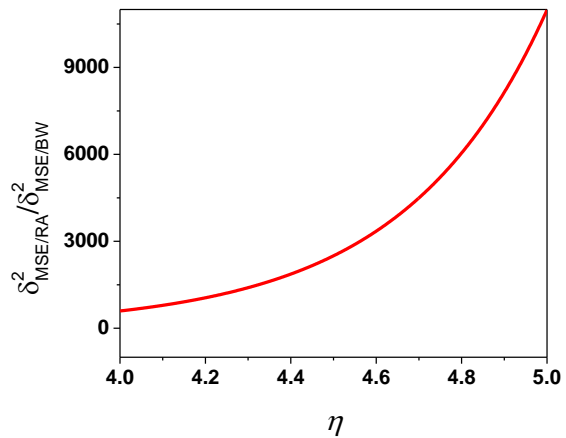

**Fig. 3.** MSE ratio $\delta^2_{MSE,RA}/\delta^2_{MSE,BW}$ of the RA filter for $\eta \leq 1.5$. If the ratio is greater than 1, as seen from $0 \leq \eta \leq 0.2$ and $\eta > 0.97$, the BW filter is superior.

**Fig. 4.** As Fig. 3, but for a more typical range of $\eta$ used in practice. Here, the BW filter is superior by orders of magnitude, with only a 10% penalty in rms noise.

## 4. Gauss-Hermite and cosine-terminated filters

In this section we consider the Gauss-Hermite (GH) [7] or extended-Gauss (EG) [9] and cosine-terminated (CT) filters as alternatives to the BW filter. The CT filter is developed here. Both offer better performance than the BW filter regarding the MSE, but approach filter design from opposite directions. The GH filter, introduced by Hoffman *et al.*, [7] follows the standard Butterworth recipe [25] of eliminating low-order coefficients in the Taylor-series expansion of the transfer function about $k = 0$. The result can be described mathematically either in terms of a product of $e^{-\eta^2}$ with an $M^{\text{th}}$-order partial series expansion of $e^{\eta^2}$ [9], or equivalently, as an incomplete gamma function [7]. It can also be represented as an $M^{th}$-order differentiation of $e^{-x^2/\Gamma^2}$ with respect to $\Gamma$ [9].

The CT is based on a different strategy, working from the cutoff end. $B(k)$ remains equal to 1 up to a value $k = k_1$, then decreases to zero at $k = k_2$, slowly at first and more rapidly at the end to take advantage of the exponential decrease of $|F(k)|^2$ to minimize overall errors. Special cases using less favorable parameters include the Tukey [28] and Hann [29] filters, and an approximation of the Welch filter [30], as noted below. While other options are available [31], these tend to exhibit greater spread and hence greater MSEs. Basing a filter on the cosine function better matches the mathematics to Fourier analysis. The result not only improves filtering, but also significantly increases computation speeds.

These filters are compared as described above. The relevant scaling factors are adjusted so $b(x_o)/b(0) = 0.5$, so the DS cutoffs remain the same throughout. Noting that all expressions are functions of $\eta = \Gamma/x_o$, for simplicity we set $x_o = 1$, so $\eta = \Gamma$.

Transfer functions $B(k)$ for the GH filter for $M$ = 1, 2, 5, 10, 20, 50, and 100 are shown in Fig. 5 for $\eta = 3$. For large values of $M$, $B(k)$ is essentially symmetric about the BW cutoff, to which it reduces in the limit $M \to \infty$. The corresponding ratios $MSE_{EG}/MSE_{BW}$ are shown as a function of $\eta$ in Fig. 6. The GH filter is seen to be an improvement over the BW filter, although this is realized only at high values of $M$, considerably higher than the values 3 and 4 considered optimal by Hoffman et al. [7]. Lower (single-digit) values of $M$ are significantly inferior to the BW in typical filtering ranges of interest for reasons that are clear from Fig. 5 and Eq. (14e): for these values $B(k)$ decreases significantly below 1 well before cutoff. The MSE improves with increasing $M$, reaching a minimum of 0.82 at $M = 100$, where because of computational overhead we terminated the calculation.

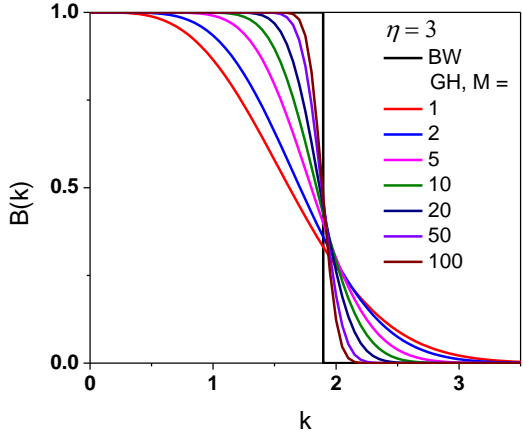 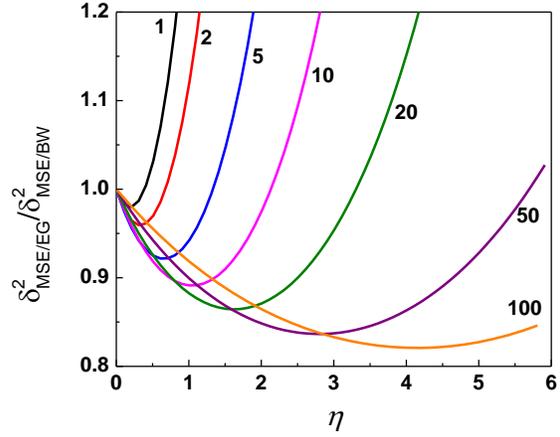

**Fig. 5.** Transfer functions of the GH filter for $M = 1, 2, 5, 10, 20, 50,$ and $100$.

**Fig. 6.** MSE ratios of the GH filter for the $M$ values of Fig. 5.

The CT filter is defined by its transfer function

$$B(k) = 1 \qquad \text{for } 0 \leq k \leq k_1; \tag{41a}$$

$$= a\cos\left(\frac{k - k_1}{\Delta k}\right) - a + 1 \quad \text{for } k_1 \leq k \leq k_2; \tag{41b}$$

$$= 0 \qquad \text{for } k > k_2. \tag{41c}$$

$k_1$, $a$, and $\Delta k$ are the parameters of the filter, and $k_2$ is the value of $k > k_1$ where $B(k_2) = 0$. $k_2$ is given by

$$k_2 = k_1 + \Delta k \cos^{-1}(1 - 1/a). \tag{42}$$

$k_1$ establishes the onset, which for comparison purposes is defined here by the requirement $b(x_o)/b(0) = 0.5$. $a$ the steepness of the cutoff, and $\Delta k$ the spread. $a$ ranges from $1/2$ to infinity. The value $a = 1/2$ yields the Tukey filter [28,31–33], characterized by a symmetric half-cycle cosine termination approximating sigmoid behavior. At $k_1 = 0$ and $a = 1$, the result approximates a Welch window, which is an inverted parabola [30]. As $a$ approaches infinity, $k_2$ becomes equal to $k_1$, and the result is the BW filter.

The general characteristics of the CT cutoffs are shown in Figs. 7a and 7b. Figure 7a illustrates the cutoff as a function of width $\Delta k$ for $a = 5$, and Fig. 7b as a function of $a$ for $\Delta k = 0.5$. Figure 7a also shows a cutoff for the EG filter with $M = 100$. Generally speaking, we find the best performance with values of $a$ of the order of 5, and $\Delta k$ of the order of 0.5. As can be seen, for these values the filter retains $B(k) = 1$ until very near cutoff, and as the cutoff begins, favors values of $B(k)$ as close to 1 as possible before dropping rapidly to zero afterward. This functional form capitalizes on the exponential decrease of $|F(k)|^2$ with increasing $k$.

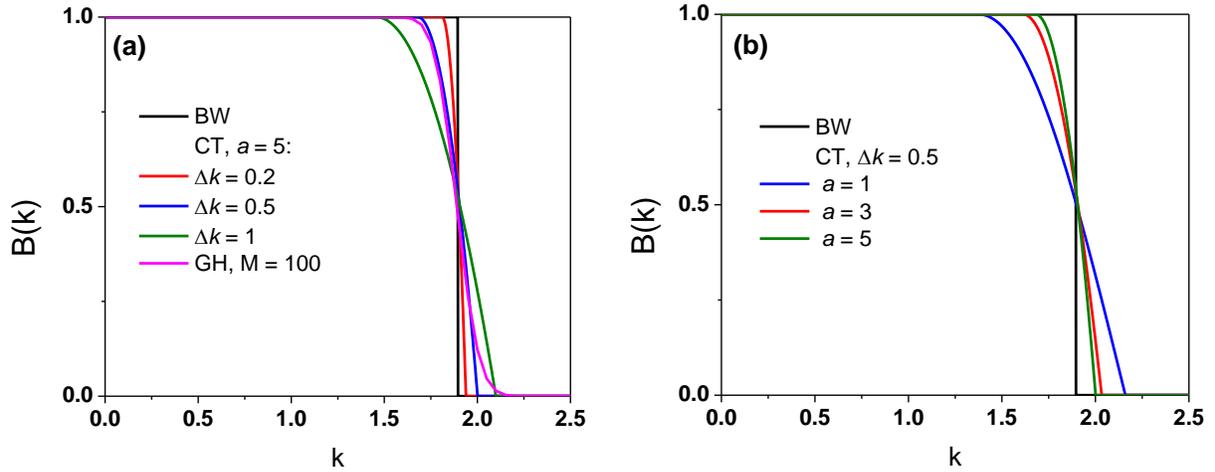

**Fig. 7.** (a) Transfer functions of the CT filter for $x_0 = 1$, $a = 5$, and $\Delta k = 0.2,\ 0.5$, and 1. The transfer functions of the BW and GH, M = 100 filters are shown for comparison. (b): As (a) but for $\Delta k = 0.5$ and $a = 1$, 3, and 5.

The Fourier transform of Eqs. (41) and (42) is

$$b(x) = b_1(x) + b_2(x) + b_3(x)$$

$$= \frac{1}{\pi x}\left(\sin k_1 x + (-a+1)(\sin k_2 x - \sin k_1 x)\right)$$

$$+ \frac{a}{2\pi(x + 1/\Delta k)}\left(\sin\left[(k_2 - k_1)(x + \frac{1}{\Delta k}) + k_1 x\right] - \sin k_1 x\right)$$

$$+ \frac{a}{2\pi(x - 1/\Delta k)}\left(\sin\left[(k_2 - k_1)(x - \frac{1}{\Delta k}) + k_1 x\right] - \sin k_1 x\right). \tag{43}$$

The apparent singularities in Eq. (43) do not exist in practice. At $x \sim 0$ the first term reduces to $(k_2 - a(k_2 - k_1))$, while at $x \sim \pm 1/\Delta k$ the others with vanishing denominators reduce to

$$b_2\left(-\frac{1}{\Delta k}\right) = b_3\left(\frac{1}{\Delta k}\right) = \frac{a(k_2 - k_1)}{2\pi} \cos k_1 x . \tag{44}$$

As noted above, the condition

$$\frac{b(x_0)}{b(0)} = \frac{1}{2} \tag{45}$$

defines $k_2$ as a function of $a$, $\Delta k$, and $k_1$ as given by Eq. (42). Taking advantage of these relations, $b_1(x)$, $b_2(x)$ and $b_3(x)$ are rewritten as

$$b_1(x) = \frac{1}{\pi x}\left(\sin k_1 x + (-a+1)\left(\sin(x(k_1 + \Delta k \cos^{-1}(1 - \frac{1}{a}))) - \sin k_1 x\right)\right) \tag{46a}$$

$$b_2(x) = \frac{a}{2\pi(x + 1/\Delta k)}\left(\sin\left(\Delta k \cos^{-1}(1 - \frac{1}{a})(x + \frac{1}{\Delta k}) + k_1 x\right) - \sin k_1 x\right) \tag{46b}$$

$$b_3(x) = \frac{a}{2\pi(x - 1/\Delta k)}\left(\sin\left(\Delta k \cos^{-1}(1 - \frac{1}{a})(x - \frac{1}{\Delta k}) + k_1 x\right) - \sin k_1 x\right). \tag{46c}$$

Numerical evaluation of these expressions shows that $b(x_0)/b(0) = 0.5$ as required by Eq. (45).

The MSE ratios for the CT filter for $a = 5$ and various $\Delta k$ are shown in Fig. 8. The CT and GH filters are compared for various $\Delta k$ and $M$ in Fig. 9, with $M$ selected to give nearly similar functional dependences on $\eta$. The results of the two filters are comparable. The Tukey filter produces results that are very close to the GH filter, as can be expected since for GH filters of large $M$ both filters have similar characteristics. The main difference between the GH and CT filters is computational speed: for comparable values of $\Delta k$ and $M$, those for the CT filter are better by at least two orders of magnitude.

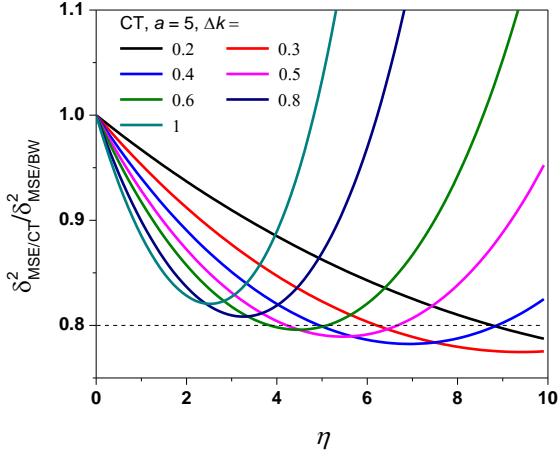 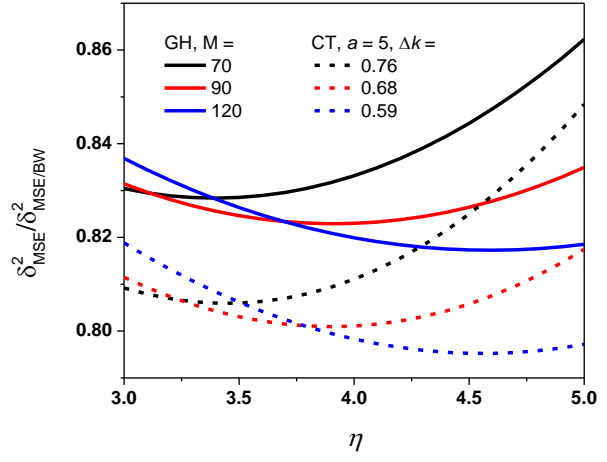

**Fig. 8.** MSE ratios of the CT filter for various $\Delta k$ for $a = 5$.

**Fig. 9.** Comparison of MSE ratios for GH and CT filters for transfer functions $B(k)$ of similar shapes. Results for the Tukey filter are essentially identical to that of the GH filter.

## 5. Discussion

As noted above, our calculations are done in the continuum limit to obtain relatively simple analytic expressions and hence to investigate linear filters without the additional complications introduced by discrete analysis. While summations can often be done in closed form, these are significantly more complicated than the continuum equivalents given above, even though they reduce properly in the continuum limit. Digital Fourier analysis also brings in the additional constraint of periodicity. This typically results in endpoint-discontinuity artifacts, although methods are now available to eliminate them [26,27].

Many of the expressions given above can be converted to their discrete equivalents for $(2N+1)$ data points by replacing $x$ with

$$x \to \theta_j = \frac{2\pi}{2N+1} j, \quad -N \le j \le N; \tag{47a}$$

$$f(\theta_j) = f_j; \tag{47b}$$

multiplying all integrals by the one-dimensional point density of states

$$\rho_{2N+1} = \frac{2N+1}{2\pi}; \tag{48}$$

and replacing all integrals from minus to plus infinity with sums from $j = -N$ to $N$. For the Fourier coefficients of the Lorentzian line defined in Eqs. (35) and (36), this transformation yields

$$f_j = \frac{1}{2N+1} \sum_{\kappa=-N}^{N} e^{-\kappa\Gamma} e^{i\kappa\theta_j} = \frac{1}{2N+1} \left( -1 + 2\text{Re} \sum_{\kappa=-N}^{N} e^{-\kappa(\Gamma - i\theta_j)} \right) \tag{49a}$$

$$= \frac{1}{2N+1} \left( -1 + 2\text{Re} \frac{1 - e^{-(N+1)(\Gamma - i\theta_j)}}{1 - e^{-(\Gamma - i\theta_j)}} \right) \tag{49b}$$

$$\cong \frac{1}{2N+1} \left( \frac{1 - e^{-2\Gamma}}{1 + e^{-2\Gamma} - 2e^{-\Gamma} \cos\theta_j} \right). \tag{49c}$$

Equation (49c) is also the maximum-entropy result for the spectrum of a single Lorentzian oscillator [34]. In the limit where $\Gamma$ and $\theta$ are small enough so second-order expansions are applicable, then

$$f_j \to \frac{1}{2N+1} \frac{1}{\Gamma^2 + \theta_j^2}, \tag{49d}$$

where Eq. (49d) is Eq. (49b) in the limit that $\Gamma$ and $\theta$ both approach zero. Thus the periodic pseudo-Lorentzian Eq. (49b) reduces to the Lorentzian in the limit of large $N$ and small $\theta$ and $\Gamma$.

Some filtering algorithms are digital only, for example the binary algorithm [12] and the entire set of weighting coefficients of Savitzky and Golay [15] and subsequent contributors [17]. The major advantage of these approaches is that they are finite, and in the SG case, involve mainly single-digit or low-double-digit numbers, an additional advantage when calculations had to be done manually. As calculations are now done by computer, this is less of a consideration. In our experience, the GH and CT filters exhibit better performance than the digital versions of the SG filters in all respects. This aspect will be covered elsewhere.

As noted above, owing to their relatively abrupt cutoffs, all high-performance linear filters generate Gibbs or cutoff oscillations when their RS coefficients are transformed to DS for use as convolution filters. An example is given in Fig. 10. From Eq. (22), the amplitude and period of these oscillations are basically determined by the last Fourier coefficient retained in the filtering process. As is also seen, making cutoffs less abrupt improves the DS convergence, but the tradeoff between rapid convergence and low MSE values remains.

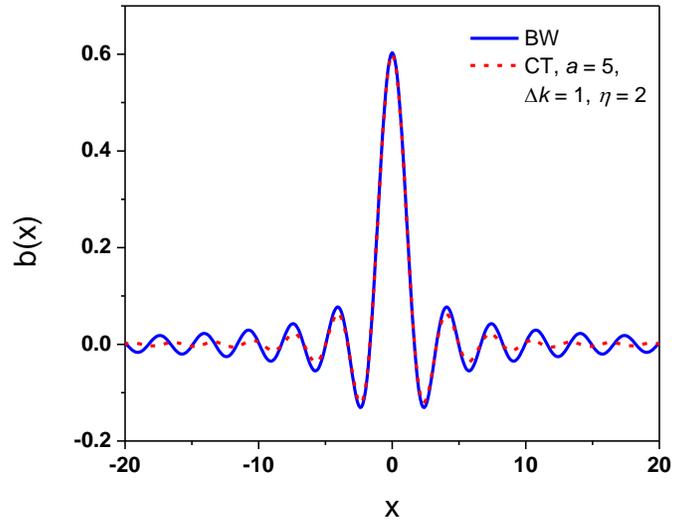

**Fig. 10.** DS lineshapes of the BW and CT filters with the same DS cutoffs. The CT parameters are shown in the figure.

Gibbs oscillations also appear in reconstructions of filtered lineshapes. An example is given in Fig. 11. As the most dominant contribution of high-order coefficients occurs when they add coherently, it is not surprising that the greatest effects are seen at extrema such as peak heights. Therefore, these are a measure of the effectiveness of a filter. Again, these artifacts can be reduced significantly or even eliminated by nonlinear methods.

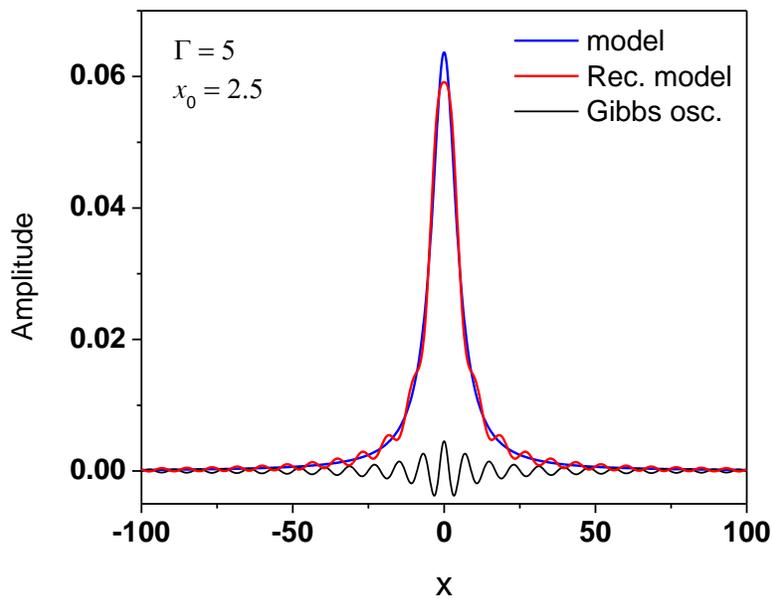

**Fig. 11.** (Black curve) Gibbs oscillations for one possible DS reconstruction of a filtered Lorentzian lineshape, shown along with the filtered and original Lorentzian lineshapes (red and blue curves, respectively). The peak amplitude of the discrepancy is proportional to $e^{-k_c \Gamma}$, and the period is approximately $\Delta x = 2\pi/k_c$. The largest difference occurs at the main peak.

## 6. Conclusion

It has long been appreciated that intelligent filtering requires that the data be examined in RS [2,3,8,9]. We find that similar advantages result if the measures that assess the filters – mean – square error and noise – are also considered in RS. We accomplish this by taking advantage of Parseval's Theorem to express these assessments in RS. The resulting perspective leads to insights that cannot be achieved from DS considerations alone, and opens up additional possibilities in linear filtering.

Using this approach, we quantitatively analyze various filters in terms of their performance using a representative Lorentzian line as data. From this perspective the RS-rectangular ("ideal" or "brick-wall") filter is significantly better than its reputation would suggest, a consequence of rigorously preserving low-order coefficients where the major fraction of spectral information resides. The BW filter can be improved by modifying its abrupt cutoff. The best previous filter that we have found to do this is the Gauss-Hermite filter, although to achieve this goal higher orders must be used than recommended in the original publication [7]. By taking advantage of the information provided by the RS formulation, we develop a cosine-terminated filter that outperforms the GH filter with a significantly reduced computation load.

We have not addressed methods that process data as well as remove noise. Processing includes operations such as interpolation, scale change, scale inversion, and differentiation. With minor modifications, the approach developed here can be used to assess performance of algorithms that do multiple operations as well.

**Acknowledgments**

This research was supported by a National Research Foundation of Korea (NRF) grant funded by the Korea government (MSIP) (NRF-2020R1A2C1009041).

## Disclosures

The authors declare no conflicts of interest.